\documentclass[iop,apj,onecolumn]{emulateapj}

\usepackage{graphicx}
\usepackage{amsmath}
\usepackage{bm}

\newcommand{\eq}{Equation~}

\newcommand{\Fig}{Figure~}

\newcommand{\fig}{Figure~}
\newcommand{\figs}{Figures~}

\newcommand{\ie}{i.e.,}
\newcommand{\eg}{{e.g.,}}
\newcommand{\cf}{{cf.}}

\newcommand{\dd}{\mathrm{d}}
\newcommand{\nablab}{\bm{\nabla}}
\newcommand{\uT}{u_{\rm T}}
\newcommand{\bu}{\boldsymbol{u}}

\begin{document}

\title{\uppercase{Transport by meridional circulations in solar-type stars}} 
\author{\sc T.~S.~Wood and N.~H.~Brummell}
\affil{Department of Applied Mathematics and Statistics, Baskin School of Engineering, University of California Santa Cruz}
\email{tsw25@soe.ucsc.edu}

\begin{abstract}
  Transport by meridional flows
  has significant consequences for stellar evolution,
  but is difficult to capture in
  global-scale
  numerical simulations
  because of the wide range of timescales involved.
  Stellar evolution models
  therefore
  usually adopt parameterizations for
  such transport
  based on idealized laminar or mean-field models.
  Unfortunately, recent attempts to model this transport in global
  simulations
  have produced results that are not consistent with
  any of these idealized models.
  In an effort
  to explain the discrepancies between global simulations and idealized models,
  we here use three-dimensional local Cartesian simulations of compressible convection to study
  the efficiency of transport by meridional flows
  below a convection zone
  in several parameter regimes
  of relevance to the Sun and solar-type stars.
  In these local simulations we are able to establish the correct \emph{ordering}
  of dynamical timescales, although the \emph{separation} of the timescales
  remains unrealistic.
  We find that, even though the generation of internal waves by convective overshoot
  produces a high degree of time dependence
  in the meridional flow field,
  the mean flow has the qualitative behavior
  predicted by laminar, ``balanced'' models.
  In particular, we observe a progressive deepening,
  or ``burrowing'', of the mean
  circulation if the local Eddington--Sweet timescale is shorter than
  the viscous diffusion timescale.
  Such burrowing is a robust prediction of laminar models in this
  parameter regime,
  but has never been observed in any previous numerical simulation.
  We argue that
  previous simulations therefore underestimate the transport by meridional flows.
\end{abstract}


\section{\uppercase{Introduction}}

\subsection{The Role of Meridional Circulations in Transport and Mixing}

The interiors of stars, in common with all rotating fluids,
frequently exhibit meridional circulations,
and the transport of angular momentum and chemical species
by these circulations has significant consequences for stellar evolution
\citep[\eg][]{Mestel53,ChaboyerZahn92,Pinsonneault97}.
Within convective zones, transport by meridional circulations is generally less
important than transport by turbulent convective motions, but within radiative
zones transport by meridional circulations may be a dominant mechanism.
Such transport may play a significant role in the spin-down of the Sun's
core \citep{Howard-etal67,Spiegel72,Clark75} and the depletion of lithium in solar-type stars
\citep[\eg][]{CharbonneauMichaud88,GaraudBodenheimer10}.
Meridional flows also transport magnetic flux, a fact that forms the
basis of certain theories of the solar magnetic cycle \cite[\eg][]{DikpatiCharbonneau99}
and the solar interior rotation \citep{GoughMcIntyre98}.

In one-dimensional (1D) models of stellar interiors, transport by meridional circulations
is sometimes parameterized
in terms of a ``rotationally-induced mixing'' parameter \citep{Zahn92}
derived from idealized laminar or mean-field models.
Many such models have been proposed, leading to many different prescriptions for transport and mixing,
all of which assume
that the amplitude and radial extent of the circulation, and hence the
degree of mixing, depend only on the
spherically-averaged angular velocity, $\Omega(r)$.
Such a description is clearly limited: in the Sun, for example,
the spherically-averaged angular velocity is almost uniform
\citep{Couvidat-etal03},
yet the \emph{latitudinal} differential rotation drives a meridional
circulation, as described in Section~\ref{sec:driving}.
Moreover, the validity of such parameterizations has never been
confirmed
by any 3D numerical model.

In this paper we
use a fully compressible, 3D numerical model
to study
the behavior of meridional flows
driven by differential rotation
of the convective envelopes of solar-type stars.
We compare the results of our simulations with the predictions of both
laminar and mean-field models and test the assumptions on which those
models are based.
In the present work we consider only non-magnetic processes;
the effects of magnetic fields will be addressed
in future papers.
Ultimately, our aim is to construct a better parameterization for the role
of meridional flows in transporting angular momentum, chemical species,
and magnetic fields in stellar interiors.

The rest of the paper is structured as follows.
In Sections~\ref{sec:driving}--\ref{sec:review} we briefly review
the physical mechanisms that give rise to
meridional flows and discuss their expected behavior
in the context of stellar interiors.
In Section~\ref{sec:global} we summarize the results of recent global
numerical simulations of angular momentum transport in the solar interior.
Our numerical model
is described
in Section~\ref{sec:model},
and parameter constraints are outlined in Section~\ref{sec:parameters}.
In Section~\ref{sec:results} we present results from four simulations
performed in different parameter regimes.
Our findings are summarized in Section~\ref{sec:summary}
and discussed in relation to the results of other studies.

\subsection{Driving of Meridional Circulations}
\label{sec:driving}

The presence
of meridional circulations in stellar interiors can
be explained by two complementary arguments.
The first
is a generalization of the classical Vogt--Eddington argument
\citep{Vogt25,Eddington25}.
In a rotating star,
a balance between centrifugal, Coriolis, pressure, and gravitational forces
in the meridional plane generally requires that temperature is non-constant
within each horizontal surface, implying that the radiative heat flux has
nonzero divergence.
Within a radiative (\ie\ non-convecting) zone,
advection of entropy by a meridional flow is the only mechanism
that can balance the divergence of the heat flux.
A meridional circulation must therefore be present in order to
maintain local thermal equilibrium and circumvent the so-called
``von Zeipel paradox'' \citep[\eg][\S5.4]{Mestel99}.

For a given internal rotation profile, we can in principal
deduce the meridional flow required to maintain the balances just described.
Although it is tempting to say that the rotation profile ``drives'' the meridional flow,
is it more correct to say that the persistence of the rotation profile, on timescales
for which meridional force balance and local thermal equilibrium are expected to apply,
\emph{implies} the presence of a meridional flow.
The transport of angular momentum by this meridional flow will
feed back onto the rotation profile,
causing it to evolve over time.
The classical example of this problem considers
a star initially in uniform rotation \citep{Sweet50};
in that case the meridional circulation that arises is known as the
Eddington--Sweet circulation.
Advection of angular momentum by this circulation causes the
star to develop differential rotation on the Eddington--Sweet timescale,
\begin{equation}
	t_{\rm ES} = \left(\frac{N}{2\Omega}\right)^{\!2} R^2/\kappa,
	\label{eq:tES}
\end{equation}
where
$\Omega$ is the initial rotation rate, $N$ is the buoyancy frequency,
$\kappa$ is the thermal diffusivity,
and $R$ is the stellar radius.
For a solar-type star,
this timescale is typically longer than the star's main-sequence lifetime, and so
transport by the Eddington--Sweet circulation is usually neglected in stellar evolution models.
However, in a differentially rotating star the meridional flows implied by the Vogt--Eddington
argument can be much stronger than the classical Eddington--Sweet circulation, particularly
in regions where the angular velocity gradient is large.  In that case transport
by the circulation can be significant.

In the classical Eddington--Sweet problem it is
assumed that the interior of the star is
not subject to any internal torques
arising from
viscous, Maxwell, or small-scale Reynolds stresses,
and so
meridional circulations are the only
means of transporting
angular momentum.
In fact, the consideration of such torques provides the
second argument
for the presence of meridional flows.
In a quasi-steady state, any torque arising from viscous, Maxwell, or Reynolds stresses
must be balanced by a Coriolis torque,
because neither pressure nor gravity has a mean azimuthal component, and
a meridional flow must be present to provide this Coriolis torque.
The process by which an applied torque drives a meridional flow
is often called ``gyroscopic pumping''
\citep{McIntyre00} by analogy with the Ekman pumping that occurs within Ekman
layers.
One example is
the solar convection zone, in which
turbulent Reynolds stresses maintain
a state of differential rotation by systematically transporting angular momentum
from high to low latitudes \citep[\eg][]{Miesch05}.
This turbulent Reynolds torque, prograde in low latitudes and retrograde in high latitudes,
also gyroscopically pumps a meridional circulation, as originally discussed by
\citet{Kippenhahn63}.
In general, part of
the circulation must extend beneath the convection zone and into the
radiation zone (unless by chance the vertically-integrated pumping torque within the
convection zone happens to be exactly zero --- see \citet[][\S8.2]{McIntyre07}).

The two arguments summarized here make slightly different assumptions about the balances
of momentum and internal energy, and so one or the other may be preferred
under different circumstances.
The gyroscopic pumping argument is in a sense more general, since it
does not assume the presence of stable stratification.
In practice, the two arguments are often complementary; for example,
the presence of meridional flows below the solar convection zone can also
be inferred from the differential rotation of the solar tachocline, using the
Vogt--Eddington argument \citep{SpiegelZahn92,GoughMcIntyre98}.

\subsection{Predictions from Laminar and Mean-Field Models}
\label{sec:review}

An important property of
gyroscopically pumped
meridional circulations is their tendency
to ``burrow'' through stably stratified regions; that is, the circulation
extends progressively deeper over time.
In this way, a meridional circulation that is gyroscopically pumped in the convective
envelope of a solar-type star can burrow into the radiative interior,
exchanging angular momentum between the two zones.

This burrowing process
was originally studied in connection with the problem of
solar spin-down, that is, the the gradual extraction of angular momentum
from the solar interior caused by magnetic braking at the Sun's surface
\citep{Schatzman62,Howard-etal67}.
Such studies usually assumed idealized, laminar conditions,
or else relied on
laboratory analogies \citep[][]{Sakurai70,BentonClark74,Clark75}.
These studies neglected any mean effects arising from waves, instabilities,
or turbulence.
The first description of the burrowing process was given by \citet{Clark73},
for linear perturbations to a state of uniform rotation in a cylinder
of stably stratified fluid.
Assuming a balance of meridional forces
(\ie\ hydrostatic and cyclostrophic balance),
as well as local thermal
equilibrium, he showed that a change in the rotation of the boundaries
of the cylinder drove a meridional circulation within a boundary layer
whose thickness grew ``hyperdiffusively'' as $t^{1/4}$.
Advection of angular momentum by the meridional circulation caused the
angular velocity perturbation imposed at the boundary to propagate into the
cylinder at the same rate.

Burrowing of meridional circulations
is now known to be a robust feature
of models in which meridional advection dominates the transport of angular momentum
\citep[\eg][]{Haynes-etal91,SpiegelZahn92,Elliott97}.
For fluids that are ``heavily stratified'', meaning that the buoyancy frequency
$N$ far exceeds the rotation rate $\Omega$, burrowing requires the presence
of a ``thermal relaxation'' mechanism that mitigates the effect of the stratification.
For this reason, in stellar interiors the rate of burrowing is dependent on the rate of
radiative diffusion, and so the burrowing process in this context is sometimes called
``radiative spreading'' \citep{SpiegelZahn92}.
Following \citet{Haynes-etal91} and \citet{McIntyre02}
we prefer to call it ``burrowing'' because of the
circulation's tendency to extend downward, rather than upward, in a fluid with a
finite density scale height.

Assuming that the results of laminar models can be applied directly to stellar interiors,
the time required for meridional circulations to burrow
across an entire stellar radiation zone,
and thereby communicate the spin-down of the surface all the way to the center,
is of the order of the Eddington--Sweet time,
$t_{\rm ES}$, given by \eq(\ref{eq:tES}),
where now
$R$ is the radius of the radiation zone.
These spin-down circulations are closely analogous to the Eddington--Sweet circulations
described in the previous section \citep{Clark75}.
Because
the timescale $t_{\rm ES}$ is typically longer than the main-sequence lifetime of a solar-type star,
spin-down by meridional circulations
is not expected to operate
all the way
to the center of the star.
In the solar interior, the Eddington--Sweet timescale
is currently $\sim 10^{12}$~years \citep[\eg][]{Gough07}, and so \citet{Clark75}
predicted that the solar rotation rate increases significantly with depth in
the radiation zone.

This picture of solar spin-down by burrowing meridional circulations
was challenged by the advent of helioseismology, which revealed that the
solar radiation zone rotates uniformly in both radius and latitude.
This uniform rotation is incompatible with any model in which angular
momentum transport occurs only by meridional advection
\citep{SpiegelZahn92}, implying that
other transport processes are operating in the solar radiation zone.
Different authors have suggested that
anisotropic turbulence
\citep{SpiegelZahn92,ChaboyerZahn92},
internal gravity waves
\citep[\eg][]{Schatzman93,Zahn-etal97},
or primordial magnetic fields
\citep{CharbonneauMacGregor93,RudigerKitchatinov97,GoughMcIntyre98}
might be responsible.
Various parameterizations for these processes have subsequently been incorporated
into one-dimensional stellar evolution models
\citep[\eg][]{CharbonnelTalon05,Eggenberger-etal05}.

\subsection{Results from Global Numerical Simulations}
\label{sec:global}

Recently, attempts have been made to
model angular momentum transport in the solar interior
using global-scale, two- and three-dimensional numerical simulations
\citep{Rogers11,Brun-etal11,Strugarek-etal11}.
These simulations include the effects of convective turbulence,
gravity waves, and magnetic fields self-consistently,
and can therefore be used to test the predictions of the laminar and
mean-field models described above.
Unfortunately, these studies
arrive at rather different conclusions, and none of them lends support to
any of the pictures of angular momentum transport just listed.
\citet{Brun-etal11} find that angular momentum transport
below
the convection zone is dominated by viscous stresses,
whereas
\citet{Rogers11} finds that transport by meridional advection and viscous
stresses cancel, with apparently no net exchange of angular momentum
between the convection and radiation zones.  \citet{Rogers11} also
finds that the presence of a magnetic field has little effect on angular
momentum transport, whereas \citet{Strugarek-etal11} find that a global-scale
interior magnetic field dominates the angular momentum transport, but
tends to produce nonuniform rotation within the radiation zone.

None of these simulations exhibit the burrowing of meridional flows
predicted by laminar models;
in fact,
such burrowing
has never been observed in any self-consistent,
three-dimensional numerical simulation.
However, the parameter regime in which burrowing is expected to occur is
rather difficult to achieve in numerical simulations.
As originally noted by \citet{Clark73}, the transport of angular momentum
by meridional circulations acts in competition with transport
by viscosity.  Since the timescale for transport by meridional
circulation is very long, of order $t_{\rm ES}$,
the transport will be dominated by viscous stresses unless
the Prandtl number is very small
(further detail is given in Section~\ref{sec:parameters}).
Recently,
\citet{GaraudBrummell08} and \citet{GaraudAA09}
have studied the penetration of meridional flows into stellar radiative
zones using laminar, axisymmetric, steady-state models.
They find that
the ratio of $t_{\rm ES}$ and the viscous diffusion timescale
plays an important role in determining both the magnitude and structure
of meridional flows.
These results have not yet been confirmed by any self-consistent,
three-dimensional numerical model.

In an attempt to understand
the discrepancies between the results of the global-scale numerical simulations,
as well as their departures from the predictions of
earlier models,
we here study in greater detail the processes that transport
angular momentum between the convective envelope and radiative interior
of solar-type stars.
We use a local Cartesian numerical model that
incorporates the nonlinear effects of convection and gravity waves, and
that allows us to study parameter regimes that are not attainable in
global-scale simulations.

\section{\uppercase{Model}}
\label{sec:model}

\begin{figure}[h!]
  \centering%
  \includegraphics[width=8cm]{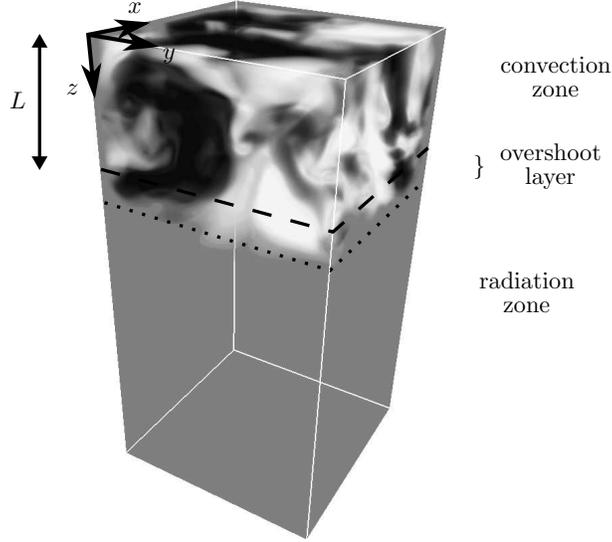}%
  \caption{%
    Illustration of the computational domain.
    The white and black coloring respectively indicates
    positive and negative values of $u_x$ at time $t=t_0$ in the simulation described
    in Section~\ref{sec:burrow}.
    The thickness of the convective layer, $L$, is fixed, but
    convective overshoot causes
    the shear flow to extend into the radiation zone.}
  \label{fig:box}
\end{figure}

We use the same code used by
\citet{Brummell-etal02}
to study the penetration
of convection into a radiation zone.
This is a fully compressible, pseudo-spectral, $f$-plane code that solves
the ideal gas equations within a Cartesian box in a rotating frame.
We adopt Cartesian
coordinates in which
$x$, $y$, and $z$ correspond to azimuth, colatitude,
and depth, respectively.
The computational domain,
illustrated in \fig\ref{fig:box},
is periodic in both horizontal directions, $x$ and $y$.
We model a localized region at the interface between the convection zone
and radiation zone.  Using a local Cartesian model, rather than
a global spherical model, has the advantage that
all of the available computational power is devoted
to studying the interaction between the two zones.
However, our results need to be interpreted in the context of the global
stellar interior.
For simplicity we take the rotation axis to be vertical
($\bm{\Omega} = -\Omega\mathbf{e}_z$),
which is a reasonable approximation at high latitudes,
where the burrowing of meridional flows is expected to
be most effective \citep[\eg][]{Haynes-etal91}.
Studies in global models \citep[\eg][]{Elliott97}
have shown that burrowing also occurs at lower latitudes,
and that the direction of burrowing is then roughly parallel to
the rotation axis.
For this reason, local and global studies of meridional circulations
tend to produce results that are qualitatively similar, but differ by a
``geometrical factor'' of order unity \citep[\eg][]{GaraudBodenheimer10}.

As in the study of \citet{Brummell-etal02},
we choose a vertical profile for the thermal conductivity, $k(z)$,
and impose a uniform vertical heat flux $H$ at the bottom of the domain.
We choose $k(z)$ and $H$ such that the radiative temperature gradient
is subadiabatic in the lower part of the domain, and superadiabatic
(\ie\ convectively unstable) in the
upper part of the domain.
This naturally leads to the formation of a lower radiation zone and
an upper convection zone.
In all the computations presented here the box size is $2L\times2L\times4L$,
where $L$ is the thickness of the convection zone.
The radiation zone is chosen to be three times thicker
than the convection zone in order to minimize
the influence of the bottom boundary on the dynamics.
The numerical resolution is typically $100\times100\times400$;
we require greater resolution in the vertical direction in order to accurately
resolve any boundary layers that form at the interface between the convection
and radiation zones.
The top and bottom boundaries of the domain
are both impenetrable and stress-free.
We impose constant temperature $T_0$ at the top of the
domain, $z=0$, and a constant heat flux at the bottom.
Initially, the fluid is at rest and in hydrostatic balance
with uniform vertical heat flux throughout,
and has pressure $p_0$ and density $\rho_0$ at the upper boundary $z=0$.
The ideal gas equations are
nondimensionalized using the thickness of the convective layer, $L$, as the lengthscale,
and $L/c$ as the timescale, where $c = \sqrt{p_0/\rho_0}$ is the isothermal sound speed
at the top of the domain.  The temperature, $T$, pressure, $p$, and density, $\rho$,
are nondimensionalized using $T_0$, $p_0$, and $\rho_0$, respectively.
The dimensionless ideal gas equations then take the form
\begin{align}
  \rho\left(\frac{\partial}{\partial t} + \bu\cdot\nablab\right)\bu
  - 2\Omega\rho\,\mathbf{e}_z\times\bu &=
  - \nablab p + g\rho\mathbf{e}_z
  + 2\mu\nablab\cdot\mathbf{D}
  + F\mathbf{e}_x
  \label{eq:mom} \\
  \left(\frac{\partial}{\partial t} + \bu\cdot\nablab\right)\rho &= -\rho\nablab\cdot\bu
  \label{eq:mass} \\
  p &= \rho T \\
  \rho T\left(\frac{\partial}{\partial t} + \bu\cdot\nablab\right)\ln(p^{1/\gamma}/\rho)
  &= \nablab\cdot\left(k(z)\nablab T\right)
  + \frac{\gamma\!-\!1}{\gamma}2\mu\mathbf{D}:\mathbf{D},
\end{align}
where $\gamma=5/3$ is the ratio of specific heats,
the constants
$\Omega$ and $g$ are the dimensionless
rotation rate and gravitational acceleration,
and $\mathbf{D}$ is the deviatoric rate-of-strain tensor,
\begin{equation}
  D_{ij} = \frac{1}{2}\frac{\partial u_i}{\partial x_j} + \frac{1}{2}\frac{\partial u_j}{\partial x_i} - \frac{1}{3}\nablab\cdot\bu\,\delta_{ij}.
\end{equation}
We have also introduced the dimensionless dynamic viscosity $\mu$ and thermal conductivity $k$,
which are related to their dimensional counterparts,
$\mu^\star$ and $k^\star$ say, by
\begin{equation}
  \mu = \frac{\mu^\star}{\rho_0cL}
  \hspace{1cm} \mbox{and} \hspace{1cm}
  k = \frac{k^\star\!/C_p}{\rho_0cL},
\end{equation}
where $C_p$ is the specific heat capacity.
We take $\mu$ to be constant throughout the domain, whereas for $k$
we impose a vertical profile
of the form
\begin{equation}
  k(z) \; = \; \frac{k_1}{1 + \exp(20(z-1))} \; + \; \frac{k_2}{1 + \exp(20(1-z))},
\end{equation}
so that $k = k_1$ in the upper layer, $z<1$,
and $k = k_2 > k_1$ in the lower layer, $z>1$,
the change occurring across a region
of dimensionless thickness $\simeq 0.1$.
The bottom of the convection zone is therefore fixed at $z=1$,
but convective motions are able to overshoot into the radiation zone.

We write the three components of the velocity field as $\bu = (u_x,u_y,u_z)$.
In our horizontally-periodic Cartesian model, differential rotation corresponds to any
$x$-averaged
flow in
the $x$ direction, and angular momentum is essentially
equivalent to
azimuthal
momentum, $\rho u_x$.
Because our computational domain is horizontally symmetric, the Reynolds stresses
in the convective layer are not able to drive any mean
differential rotation.
In order to
mimic the generation of differential rotation in the solar convection zone, we add a
volume forcing term to the $x$ component of the momentum equation (\ref{eq:mom}),
\begin{equation}
  F = \lambda(z,t)\rho(u_{\rm T}(y,z,t) - u_x).
\end{equation}
The effect of the forcing term is to push the flow toward a prescribed
``target'' shear flow $\uT$ at a rate $\lambda$.
We have chosen to consider a situation analogous to the thought-experiment of \citet{SpiegelZahn92},
in which the radiation zone is initially in uniform rotation despite the
presence of differential rotation in the convection zone.
We therefore take $\lambda$ to be constant initially, $\lambda=\lambda_0$,
and the target velocity $\uT$ is taken to be
\begin{equation}
	\uT \; = \; \frac{u_0(1-z)\sin(\pi y)}{1 + \exp(20(z-1))},
	\label{eq:uT_initial}
\end{equation}
which tends to zero at depths $z > 1$.
We thereby suppress differential rotation in the radiation zone
and drive differential rotation in the convection zone.
By suppressing differential rotation in the radiation zone
we also prevent the burrowing of meridional circulations,
which are tied to the differential rotation through the balances
described in Section~\ref{sec:driving}.

In all the computations presented here we take $\lambda_0 = 2\Omega$, so that the
forcing rate matches the rotation rate, and $u_0=2\Omega/\pi$, so that
the shearing timescale is comparable to the rotation period.
The Rossby number
$\Omega^{-1}\partial\uT/\partial y$ is therefore of order unity,
and so the differential rotation is somewhat stronger
than that observed in the Sun and
most solar-type stars \citep[\eg][]{Reiners06}.
By forcing a
strong differential rotation,
we can study the nonlinear effects
of finite Rossby number,
which were neglected in most of the idealized models reviewed in
Section~\ref{sec:review}.
Once the differential rotation reaches a statistically steady state,
at $t=t_0$ say, the forcing parameters are changed to
\begin{flalign}
  &&\lambda \; &= \; \frac{\lambda_0}{1 + \exp(20(z-1))}& \\
  &\mbox{and}& \uT \; &= \; u_0(1-z)\sin(\pi y) \hspace{2cm} \mbox{for $t>t_0$}&
\end{flalign}
so that $\lambda$ tends to zero at depths $z>1$.
This means that the suppressive forcing is switched off below the convection zone
for $t > t_0$,
allowing the differential rotation to propagate into the interior,
if the dynamics so dictate.

\section{\uppercase{Choice of parameters}}
\label{sec:parameters}

The burrowing described in Section~\ref{sec:review}
is expected to occur only under specific
parameter conditions, which are difficult to achieve in a computational model.
The relevant parameter regime
can be characterized as a hierarchy of timescales
in the radiation zone:
\begin{center}
\begin{tabular}{ccccccccc}
acoustic time & $\ll$ & buoyancy time & $\ll$ & rotation time & $\ll$ & Eddington--Sweet time & $\ll$ & viscous time
\end{tabular}
\end{center}
\citep[\eg][]{Clark73}.
In terms of our dimensionless parameters,
these conditions
can be expressed as
\begin{equation}
	1 \;\; \ll \;\; \frac{1}{N} \;\; \ll \;\; \frac{1}{2\Omega} \;\; \ll \;\;
	\left(\frac{N}{2\Omega}\right)^{\!2}\frac{W^2}{k_2/\rho} \;\; \ll \;\; \frac{W^2}{\mu/\rho}
	\label{eq:ordering}
\end{equation}
where $N$ is the dimensionless buoyancy frequency,
\begin{equation}
	N^2 = \frac{g^2}{T}\left(
	  \frac{\gamma\!-\!1}{\gamma} - \frac{1}{g}\frac{\dd T}{\dd z}
	\right),
\end{equation}
and $W$ is a characteristic horizontal lengthscale,
which
is of order unity
in our Cartesian model.
We note in particular that the viscous time
(\ie\ the timescale for viscous diffusion across the domain)
must exceed the Eddington--Sweet time.
This condition can be expressed as a constraint on the Prandtl number
$\mu/k_2$ within the radiation zone:
\begin{equation}
  \mu/k_2 \;\; \ll \;\; \left(\frac{2\Omega}{N}\right)^{\!2}.
  \label{eq:Prandtl}
\end{equation}
This constraint is particularly stringent because
the second inequality in \eq(\ref{eq:ordering}) implies that
the right-hand side of \eq(\ref{eq:Prandtl}) is $\ll 1$.
Following \citet{GaraudAA09} we introduce the dimensionless parameter
\begin{equation}
  \sigma = \frac{N}{2\Omega}\sqrt{\dfrac{\mu}{k_2}}
  \label{eq:sigma}
\end{equation}
in order to express condition (\ref{eq:Prandtl}) more succinctly as $\sigma \ll 1$.
If this condition is not met then viscous transport of angular momentum
dominates the transport by meridional flows \citep[][\S3.4]{Clark73,BentonClark74},
and we expect the burrowing of the meridional circulation to be less efficient.
For example, in the laminar, steady-state model of \citet{GaraudAA09},
cases with $\sigma > 1$ have meridional circulations that
decay exponentially beneath the convection zone, across a boundary layer
similar to that described by \citet{BarcilonPedlosky67}
\citep[see also][]{Lineykin55}.

For any solar-type star
at the start of the main-sequence,
condition (\ref{eq:Prandtl})
holds
throughout the radiation zone,
but as the star spins down through magnetic braking,
this condition may cease to hold in the most strongly stably stratified regions.
In the solar interior, for example, condition (\ref{eq:Prandtl})
holds
only within the
outer part of the radiation zone \citep{GaraudAA09}
and a very small region at the center.

Computational limitations make it difficult to achieve the
``low-sigma'' regime described by condition (\ref{eq:Prandtl}) in
a numerical model.
If the rotation rate $\Omega$ and
buoyancy frequency $N$ take realistic stellar values,
then the Prandtl number $\mu/k_2$ must be extremely small,
which requires very high spatial resolution.
For this reason,
all global simulations of the solar interior have been performed in the
``high-sigma'' regime, $\sigma>1$ \citep[\eg][]{Rogers11,Brun-etal11,Strugarek-etal11}.
It is therefore important to test whether the behavior of meridional flows
in nonlinear numerical simulations depends on $\sigma$ in the same manner
as is
predicted by laminar models.
In order to reach a parameter regime with $\sigma < 1$,
we have chosen here to use a local model, and to adopt more modest
values for $\Omega$ and $N$, while still preserving the ordering of timescales
given by (\ref{eq:ordering}).

The code of \citet{Brummell-etal02}
requires the specification of six dimensionless input parameters,
which we take to be:\vspace{-0.15cm}
\begin{itemize}
  \item the values $k_1$ and $k_2$ of the thermal conductivity in the convection and radiation zones;
  \vspace{-0.15cm}
  \item the dynamic viscosity $\mu$;
  \vspace{-0.15cm}
  \item the gravitational acceleration $g$;
  \vspace{-0.15cm}
  \item the rotation rate $\Omega$;
  \vspace{-0.15cm}
  \item the heat flux $H = k_2\partial T/\partial z$ at the bottom of the domain.
\end{itemize}\vspace{-0.15cm}
For any particular choices of $H$ and $k_1$ we can choose values for the remaining parameters
in order to satisfy the four constraints in (\ref{eq:ordering});
different choices for $H$ and $k_1$ correspond to
different degrees of compressibility and convective overshoot.\footnote{
  \citeauthor{Brummell-etal02}\ measure the degree of compressibility and
  overshoot in terms of the quantities $\theta=H/k_1$ and
  $S=
  \left.\left(\frac{gk_2}{H}-\frac{\gamma}{\gamma-1}\right)
  \middle/
  \left(\frac{\gamma}{\gamma-1}-\frac{gk_1}{H}\right)\right.$.
  All the convection simulations presented here have $\theta=0.1$ and $S=15$.
}
In practice, we choose values for $H$ and $k_1$
such that the pressure scale height is comparable to the height of the domain,
and such that convective overshoot extends a distance of order $0.1$
below the base of the convection zone (see \fig\ref{fig:box}),
as is thought to be typical
for solar-type stars \citep[\eg][]{Brummell-etal02,Rempel04}.

\section{\uppercase{Results}}
\label{sec:results}

In the following sections we present results from four simulations performed
in different parameter regimes.
The first simulation, Case 0, differs from the others in
that the conductivity of the upper layer, $k_1$, was chosen to make that layer
adiabatic, and hence
marginally stable to convection.
The flow
in this simulation
therefore remains laminar, with no convection or internal wave
generation, and so we expect the results to follow the predictions of the
laminar models summarized in Section~\ref{sec:review}.
Case 1 has the same parameters as Case 0 for the lower layer, but has a smaller
conductivity $k_1$ in the upper layer, so that the upper layer is convectively
unstable.
Both Case 0 and Case 1
obey the ordering of timescales in (\ref{eq:ordering}), and so both have
$\sigma < 1$;
we will refer to this as the ``low-sigma regime''.

Case 2 and Case 3 both have a larger viscosity $\mu$
and smaller thermal conductivity $k$ throughout
the domain, relative to Case 1, such that $\sigma > 1$;
we will refer to this as the ``high-sigma regime''.
Case 2 has a larger viscosity and conductivity than Case 3, but both have
the same Prandtl number $\mu/k$, equal to 0.24 in the lower layer, $z>1$.
In both cases the heat flux $H$ was chosen so that the stratification
profile, and hence the buoyancy frequency $N$, matches that of Cases 0 and 1.
Case 2 is only weakly convective, because of its relatively high viscosity,
whereas Case 3 is more strongly convective, because of its low
thermal conductivity.

In each simulation the dimensionless rotation rate and gravitational acceleration
were taken to be $\Omega = 9.6\times10^{-3}$ and $g = 0.24$ respectively.
The other parameters, and relevant timescales, are listed in Table~\ref{tab:parameters}.
For reference, Table~\ref{tab:parameters} also lists characteristic values for the same
dimensionless timescales in the solar interior, based on
the values at $0.7R_\odot$ reported by \citet{Gough07}.
To allow the most direct comparison with the numerical simulations, the solar timescales
are calculated using the lengthscale $W = 0.7R_\odot \simeq 4.9\times10^{10}$\,cm,
and quoted in units of $L/c$,
where $L = 1.4\times10^9$\,cm is one quarter of the pressure scale height
and $c=\sqrt{p/\rho} \simeq 1.8\times10^7$\,cm\,s$^{-1}$ is the isothermal sound speed.
\begin{deluxetable*}{cccccccccc}

\tablecaption{Parameters of each simulation, and corresponding solar parameters}

\tablehead{\colhead{Case} & \colhead{$10^4k_1$} & \colhead{$10^4k_2$} & \colhead{$10^4\mu$} & \colhead{$10^4H$} & \colhead{$\dfrac{1}{N}$} & \colhead{$\dfrac{1}{2\Omega}$} & \colhead{$\left(\dfrac{N}{2\Omega}\right)^{\!2}\dfrac{W^2}{k_2/\rho}$} & \colhead{$\dfrac{W^2}{\mu/\rho}$} & \colhead{$\sigma$} }

\startdata
0 & 15.1 & 24.1 & 0.145 & 1.4 & 11 & 52 & 1.0$\times10^4$ & 7.9$\times10^4$ & 0.36 \\
1 & 14.5 & 24.1 & 0.145 & 1.4 & 11 & 52 & 1.0$\times10^4$ & 7.9$\times10^4$ & 0.36 \\
2 & 5.12 & 8.53 & 2.05 & 0.51 & 11 & 52 & 2.9$\times10^4$ & 0.56$\times10^4$ & 2.27 \\
3 & 1.93 & 3.22 & 0.774 & 0.19 & 11 & 52 & 7.6$\times10^4$ & 1.5$\times10^4$ & 2.27 \\
Sun & \nodata & \nodata & \nodata & \nodata & 16 & 2400 & 4.8$\times10^{16}$ & 1.1$\times10^{18}$ & 0.21
\enddata

\tablecomments{The last five columns give approximate values for the dimensionless
timescales in \eq(\ref{eq:ordering}) and the value of $\sigma$,
calculated using
the time-averaged density $\rho$ and buoyancy frequency $N$ below the overshoot region.}

\label{tab:parameters}
\end{deluxetable*}

\subsection{Case 0: The Low-Sigma Regime, without Convection}
\label{sec:laminar}

\begin{figure}[h!]
  \centering%
  \includegraphics[height=7cm]{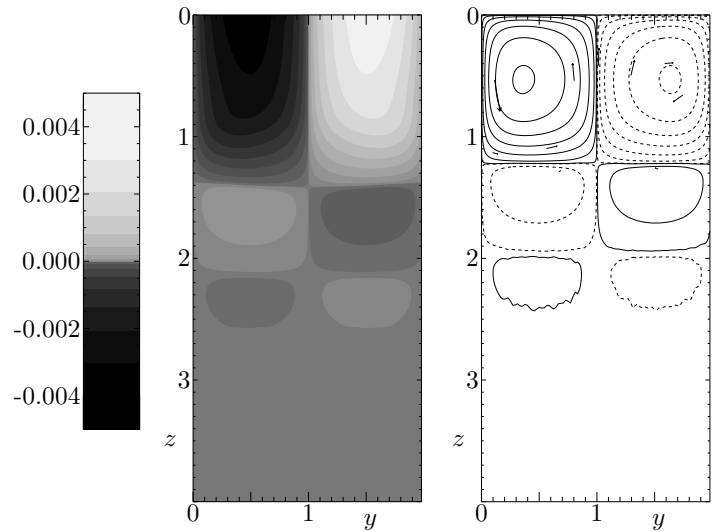}%
  \caption{%
    Shear flow $u_x$ (left panel)
    and meridional streamlines (right panel)
    from Case 0,
    averaged in $x$ at time $t=t_0$.
    The meridional streamlines are drawn as contours of a streamfunction $\psi$,
    which was computed assuming that $\nablab\cdot\rho\bu=0$.
    Solid and dashed contours indicate anti-clockwise and clockwise circulation
    respectively.
    Contour levels for $u_x$ and $\psi$ are cubically spaced to show more detail
    in the radiation zone, where the flows are weakest.}
  \label{fig:SZ-lam-t0}
\end{figure}

In this case the upper layer, $0<z<1$, is non-convective, and so we expect the
dynamics to follow the predictions of the laminar models discussed in Section~\ref{sec:review}.
\Fig\ref{fig:SZ-lam-t0} shows the steady azimuthal shear and meridional flow
at $t=t_0$.
By this time the flow
has reached a steady state
with a large-scale azimuthal shear and meridional circulation in the upper layer.
Both the shear and circulation extend slightly into the
lower layer, to an extent that depends on the rate at which the target velocity $\uT$
tends to zero for $z>1$, but for $z>1.3$ the flow is exponentially weak.
We note that the azimuthal shear $u_x$ does not exactly match the target shear flow $\uT$
given by \eq(\ref{eq:uT_initial}).  In the steady state, the ``residual'' forcing
$\lambda_0\rho(\uT-u_x)$ is balanced primarily by the azimuthal component of the
Coriolis force, $2\Omega\rho u_y$, and this balance determines the strength of
the meridional flow within the upper layer.  This is an example of the process
referred to as gyroscopic pumping in Section~\ref{sec:driving}.
Within the upper layer, the downwelling portion of the meridional circulation
is stronger than the upwelling portion.  This asymmetry is a symptom of the Rossby
number being of order unity, and is therefore less apparent in the lower layer, where the shear is weaker.

\begin{figure}[h!]
  \centering%
  \includegraphics[width=14cm]{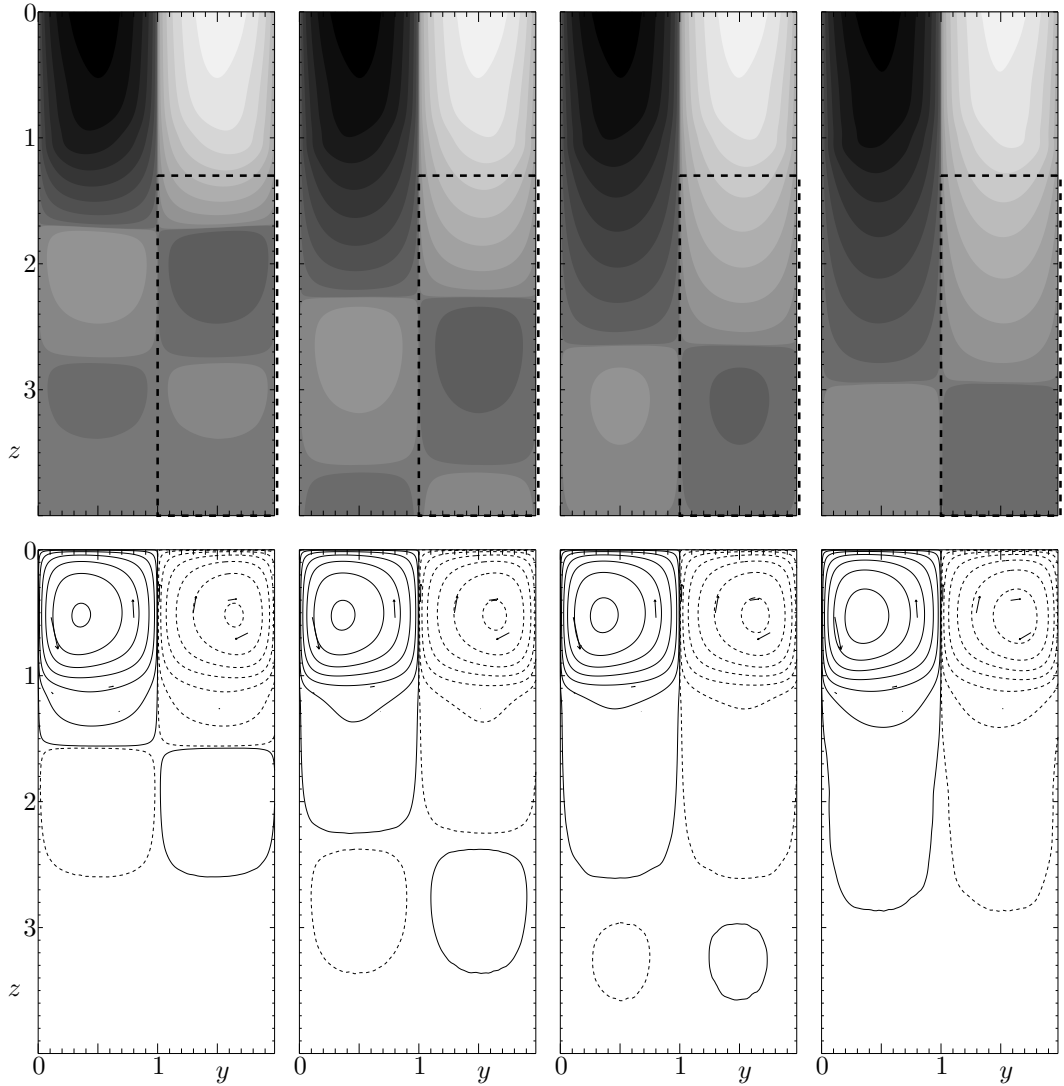}%
  \caption{%
    Shear flow and meridional streamlines from Case 0
    averaged over one rotation period for successive times $t>t_0$,
    using the same contour levels as in \fig\ref{fig:SZ-lam-t0}.
    The time averages were taken over the intervals $\Delta t_1$,
    $\Delta t_2$, $\Delta t_3$, and $\Delta t_4$ indicated in
    \fig\ref{fig:SZ-mom-lam}.
    The dashed box in the upper panels
    indicates the control volume $V$ used in \eq(\ref{eq:torque}).}
  \label{fig:SZ-lam-ave}
\end{figure}

\begin{figure}[h!]
  \centering%
  \includegraphics[width=13cm]{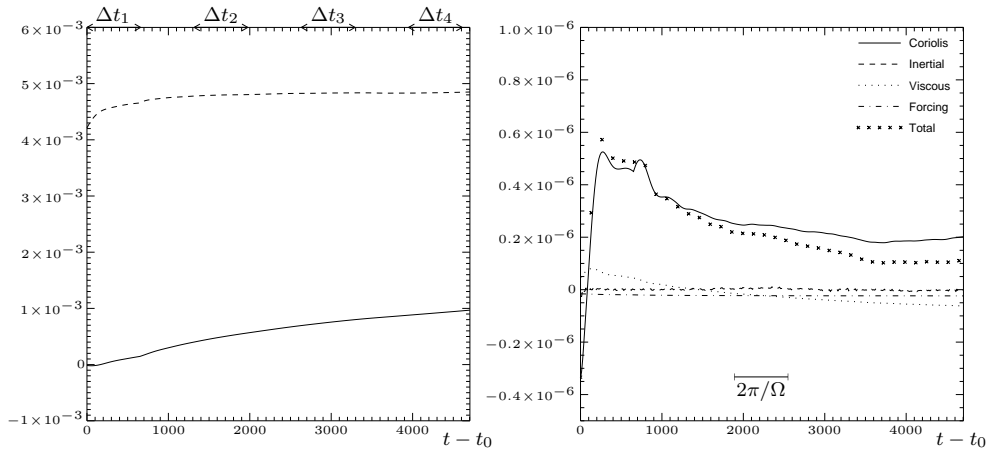}%
  \caption{%
    Left panel: The solid line shows the total azimuthal momentum $\int\dd V\rho u_x$
    in the dashed box indicated in \fig\ref{fig:SZ-lam-ave}, averaged over one
    rotation period.
    For comparison, the dashed line shows the total azimuthal momentum in the region
    \emph{above} the dashed box.
    Right panel: The four terms contributing to the right-hand side of \eq(\ref{eq:torque}),
    and their total,
    each averaged over one rotation period.
    The length of one rotation period $2\pi/\Omega$ is indicated on the plot.
    The total duration plotted corresponds to about half of the Eddington--Sweet time.}
  \label{fig:SZ-mom-lam}
\end{figure}

At $t=t_0$ the forcing is switched off below $z=1$,
allowing the flow of the upper layer to propagate into the lower, stably stratified
layer.
The sudden imbalance
of forces also generates a spectrum of internal waves.
\Fig\ref{fig:SZ-lam-ave} shows time averages of the azimuthal shear and meridional
flow, averaged over one rotation period, at regular intervals after $t=t_0$.
The time averaging filters out
most of
the internal wave modes, making the long-time
evolution more visible.
From $t=t_0$ onwards, the meridional circulation
of
the upper layer begins to burrow into the lower layer,
as indicated by the increased vertical extent of the main meridional circulation cell
in the lower panel of \fig\ref{fig:SZ-lam-ave}.
The mean azimuthal shear also propagates into the lower layer,
at approximately
the same rate,
suggesting that the transport of angular momentum is the result of advection by the
meridional flow, as expected for a laminar flow at these parameter values.
To verify this,
we first
use the mass conservation equation (\ref{eq:mass})
to write the azimuthal component of the momentum equation (\ref{eq:mom}) as
\begin{equation}
  \frac{\partial}{\partial t}(\rho u_x)
  =
  - 2\Omega\rho u_y
  -\nablab\cdot(\rho u_x\,\bu)
  + \mu\nabla^2 u_x
  + F.
  \label{eq:point_torque}
\end{equation}
After integrating
over a fixed volume $V$, and employing the divergence theorem, we obtain
\begin{equation}
  \frac{\dd}{\dd t}\int_V\!\dd V\rho u_x
  \;=\;
  \underbrace{-\;\; 2\Omega\int_V\!\dd V\rho u_y}_{\rm Coriolis}\;
  \underbrace{-\; \int_{\partial V}\!\dd\mathbf{S}\cdot\bu\,\rho u_x}_{\rm inertial}\;
  \underbrace{+\;\; \mu\int_{\partial V}\!\dd\mathbf{S}\cdot\nablab u_x}_{\rm viscous}\;
  \underbrace{+\; \int_V\!\dd V F}_{\rm forcing},
  \label{eq:torque}
\end{equation}
where $\partial V$ represents the boundary of $V$, and where $\dd\mathbf{S}$ is the
area element directed outward.
In order to verify that the burrowing meridional circulation seen in the lower panel of \fig\ref{fig:SZ-lam-ave}
is responsible for the propagation of the shear flow into the interior,
we have computed each integral in \eq(\ref{eq:torque})
for the control volume $V$ indicated by the dashed box in the upper panel of \fig\ref{fig:SZ-lam-ave}.
The result, after taking a time average over one rotation period, is plotted in
\fig\ref{fig:SZ-mom-lam}.
The left panel of \fig\ref{fig:SZ-mom-lam} shows that,
whereas the azimuthal momentum in the upper layer adjusts to a new equilibrium
after only a few rotation periods, the lower layer
adjusts on a much longer timescale.
The right panel shows that the long-time transport of azimuthal momentum
into the lower layer
can be attributed almost entirely to the
azimuthal Coriolis force arising from the mean meridional flow,
which corresponds to advection of angular momentum viewed in our rotating frame.

\begin{figure}[h!]
  \centering%
  \includegraphics[height=6cm]{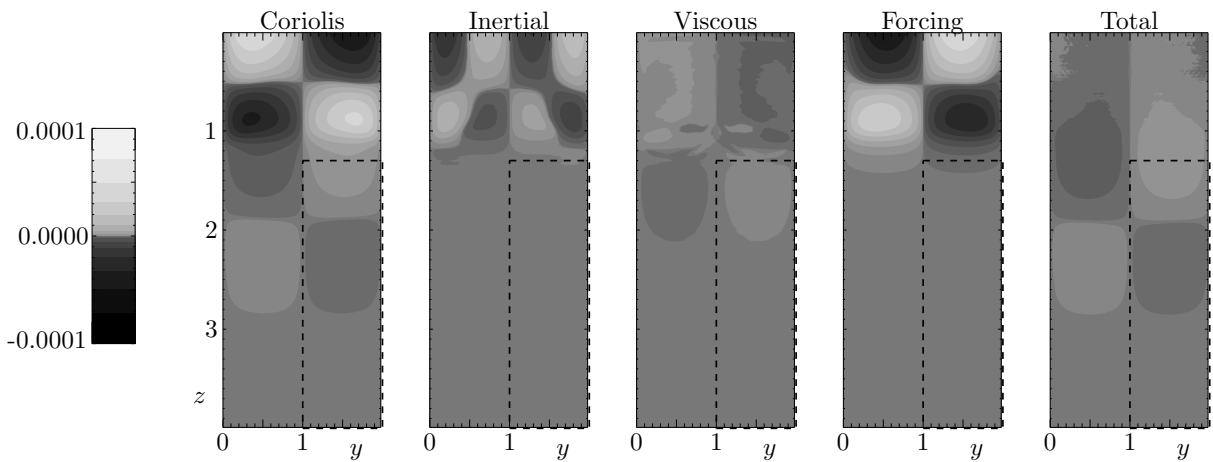}%
  \caption{%
    Vertical cross-sections of each term on the right-hand side of \eq(\ref{eq:point_torque}),
    and their total, averaged in azimuth and over the time interval $\Delta t_1$ indicated
    in \fig\ref{fig:SZ-mom-lam}.  The contour levels are cubically spaced, as indicated by
    the colorbar on the left.  Throughout the control volume $V$, indicated by the
    dashed box, the Coriolis term provides the dominant contribution to the total
    angular momentum transport.}
  \label{fig:sasha-SZ-lam}
\end{figure}

As a further illustration of the relative contributions of the different processes
to the net transport of angular momentum, in \fig\ref{fig:sasha-SZ-lam}
we present vertical cross-sections of each term on the
right-hand side of \eq(\ref{eq:point_torque}), after averaging in azimuth and over
the time interval $\Delta t_1$ indicated in \fig\ref{fig:SZ-mom-lam}.
Within the upper layer, the Coriolis and forcing terms are approximately in balance,
and the strength of the meridional flow is determined by gyroscopic pumping.
Within the lower layer, the Coriolis term dominates all others, leading to the evolution
of the differential rotation seen in \fig\ref{fig:SZ-lam-ave}.
Within this layer, the strength of the meridional flow is determined by
meridional force balance and local thermal equilibrium,
as anticipated by the Vogt--Eddington argument of Section~\ref{sec:driving}, and
is therefore roughly proportional to the vertical gradient of the angular velocity,
\ie\ to $\partial u_x/\partial z$ in our Cartesian geometry.
This explains why the differential rotation and meridional circulation propagate at the same
rate, and also why the strength of the meridional flow, and hence the
rate of burrowing, decays over time.
The results in this case are entirely consistent with the laminar models
described in Section~\ref{sec:review}.

\subsection{Case 1: The Low-Sigma Regime, with Convection}
\label{sec:burrow}

We now study the effects of turbulent convective overshoot and internal waves
on the burrowing of meridional flows.
In Case 1,
unlike Case 0,
the upper layer, $0<z<1$, is convectively unstable,
and has an rms Reynolds number $\simeq550$.
The turbulent
motions in this layer continually generate internal waves, which propagate into
the lower, stably stratified layer.
Nevertheless,
for times $t\leqslant t_0$
the time-averaged shear and meridional circulation
are confined to the upper layer,
as in Case 0.
Motions in the lower layer, $z>1$, are suppressed by the forcing in that region.

\begin{figure}[h!]
  \centering%
  \includegraphics[width=14cm]{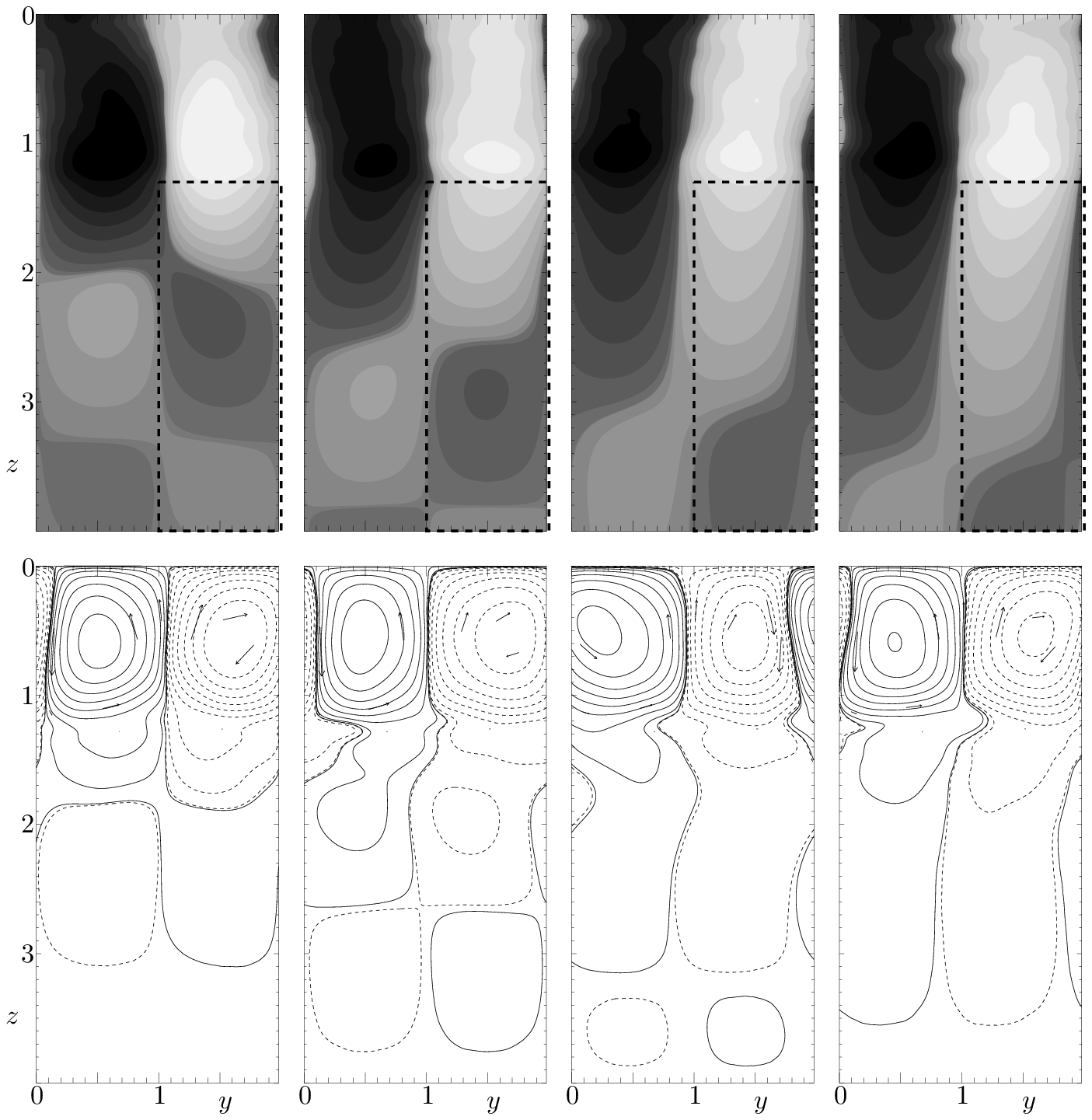}%
  \caption{%
    Shear flow and meridional streamlines from Case 1,
    averaged over two rotation periods for times $t>t_0$.
    Time averages were taken over the intervals indicated in \fig\ref{fig:SZ-mom}.
    The contour levels are the same as in \fig\ref{fig:SZ-lam-ave}.}
  \label{fig:SZ-ave}
\end{figure}

At $t=t_0$ the forcing is switched off in the lower layer.
\Fig\ref{fig:SZ-ave} shows
the mean shear and meridional flow,
averaged over two rotation periods,
at regular intervals after $t=t_0$,
using the same contour levels as \fig\ref{fig:SZ-lam-ave}.
As in Case 0,
the shear and meridional circulation both extend progressively deeper into
the lower layer over time, though in a much less regular
fashion than in Case 0.
In any snapshot of the meridional flow, the mean circulation is completely
dominated by internal wave motions,
but
after averaging over two rotation periods
we observe
a similar mean circulation to that in \fig\ref{fig:SZ-lam-ave}.

\begin{figure}[h!]
  \centering%
  \includegraphics[width=13cm]{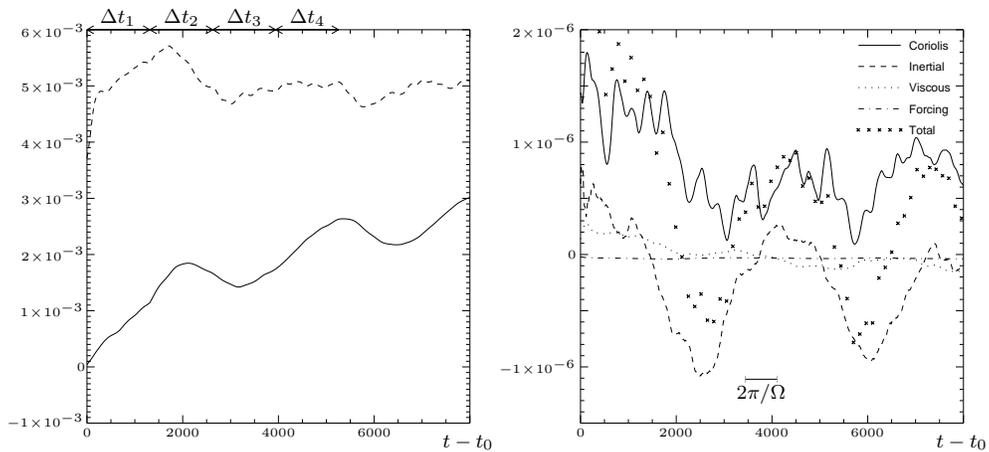}%
  \caption{%
    Same plots as \fig\ref{fig:SZ-mom-lam}, but for Case 1 and averaged over
    two rotation periods.
    The scale in the right-hand panel is larger than in \fig\ref{fig:SZ-mom-lam}.
    The inertial contributions to the momentum transport, which are associated with
    waves and turbulent fluctuations, are larger than in Case 0,
    but the Coriolis force from the mean meridional flow remains dominant
    in the long term, leading to the overall increase in azimuthal momentum
    shown by the solid line in the left panel.
    The dimensionless Eddington--Sweet time in this case is $\simeq10^4$.}
  \label{fig:SZ-mom}
\end{figure}

In \fig\ref{fig:SZ-mom} we show the evolution of the azimuthal momentum
in the
volume indicated by the
dashed box in \fig\ref{fig:SZ-ave}, as well as the
terms contributing to the
right-hand side of \eq(\ref{eq:torque}) for this volume.
(This figure is equivalent to \fig\ref{fig:SZ-mom-lam} in Section~\ref{sec:laminar}.)
Despite the presence of
internal waves, we find that the
mean meridional circulation dominates the long-time transport of azimuthal
momentum into the radiative layer, as in Case 0 for which the flow was laminar.

\begin{figure}[h!]
  \centering%
  \includegraphics[height=6cm]{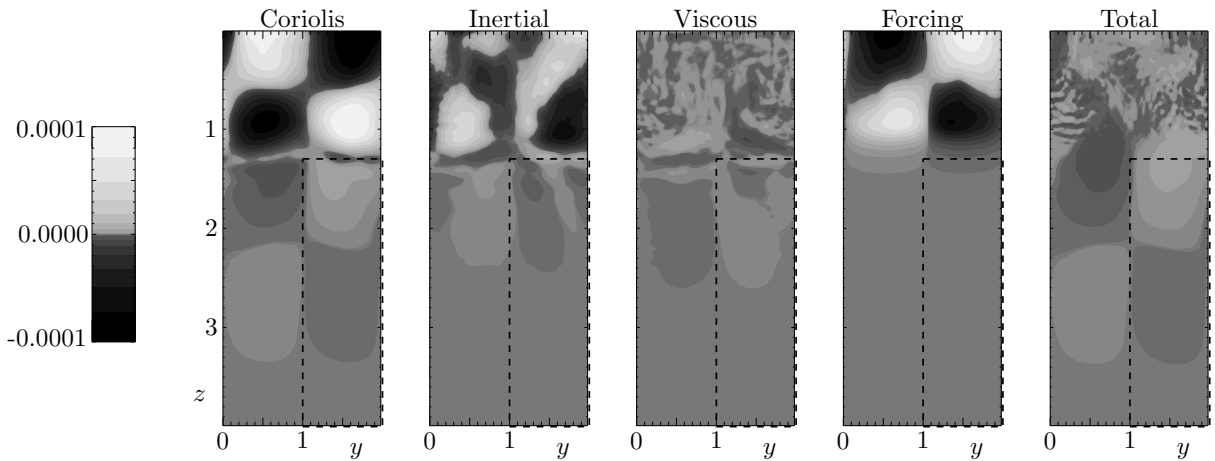}%
  \caption{%
    Same plots as \fig\ref{fig:sasha-SZ-lam}, but for Case 1
    and averaged over the time interval $\Delta t_1$ indicated in \fig\ref{fig:SZ-mom}.
    As in Case 0, the Coriolis term provides the dominant contribution to the total
    angular momentum transport in the radiation zone.}
  \label{fig:sasha-SZ}
\end{figure}

Although there is qualitative agreement between the results of Case 0
and Case 1, we note that they differ in certain respects.
In particular, the propagation of the shear into the lower layer
does not proceed monotonically, as can be seen in the left panel
of \fig\ref{fig:SZ-mom} (\cf\ \fig\ref{fig:SZ-mom-lam}),
indicating that the burrowing process does not act continuously
throughout the simulation.
Perhaps more surprising, however, is that
the overall rate of burrowing is \emph{faster} in Case 1 than in Case 0.
These differences between Case 0 and Case 1
can be attributed to differences in the
profile of $u_x$ within the upper layer in the two cases.
In Case 0, $u_x$ is strongest close to the
upper boundary, $z=0$, where the forcing is strongest.
In Case 1, $u_x$ is
highly time-dependent, but
generally
strongest close to the interface $z=1$ (see \fig\ref{fig:SZ-ave}).
Therefore the amplitude of $u_x$
at the top of the radiation zone is generally stronger in Case 1,
but varies significantly in time,
producing a burrowing that is more rapid
overall, but very irregular.
The occasional reversals in the sign of the momentum transport in \fig\ref{fig:SZ-mom}
(\eg\ for $2000\leqslant t \leqslant 3000$)
correspond to the periods when $u_x$ at the interface is weakest.
The variations in $u_x$ in the upper layer are quasi-periodic, with a timescale of several rotation
periods, and may possibly be the result of an inertial mode trapped within the convection zone.
(An inertia--gravity oscillation within the radiation zone, on the other hand, would necessarily have a period of less than half the rotation period, because $N > 2\Omega$.)
In that case the oscillation relies on the artificial local geometry of our model, and would not
be present in a global model.
Finally, we note that during the phases when burrowing does occur, the transport of
angular momentum throughout the radiation zone is qualitatively similar to that in Case 0.
This is illustrated in \fig\ref{fig:sasha-SZ}, in which we plot vertical cross-sections of each term
on the right-hand side of \eq(\ref{eq:point_torque}), averaged in azimuth and over the interval
$\Delta t_1$ indicated in \fig\ref{fig:SZ-mom}.
(This figure is equivalent to \fig\ref{fig:sasha-SZ-lam} in Section~\ref{sec:laminar}.)
Although there is a non-zero contribution from the inertial and viscous terms,
the main contribution comes from the Coriolis term,
and the overall transport of angular momentum in the radiation zone resembles that in Case 0.

\subsection{Case 2: The High-Sigma Regime; High Viscosity}

In Case 2 the viscous diffusion timescale is shorter than the
Eddington--Sweet timescale, primarily because the viscosity is
larger than in Cases 0 and 1, and so $\sigma > 1$.
The rotation rate $\Omega$ and stratification profile
in the radiation zone
are the same as in Cases 0 and 1.

\begin{figure}[h!]
  \centering%
  \includegraphics[width=14cm]{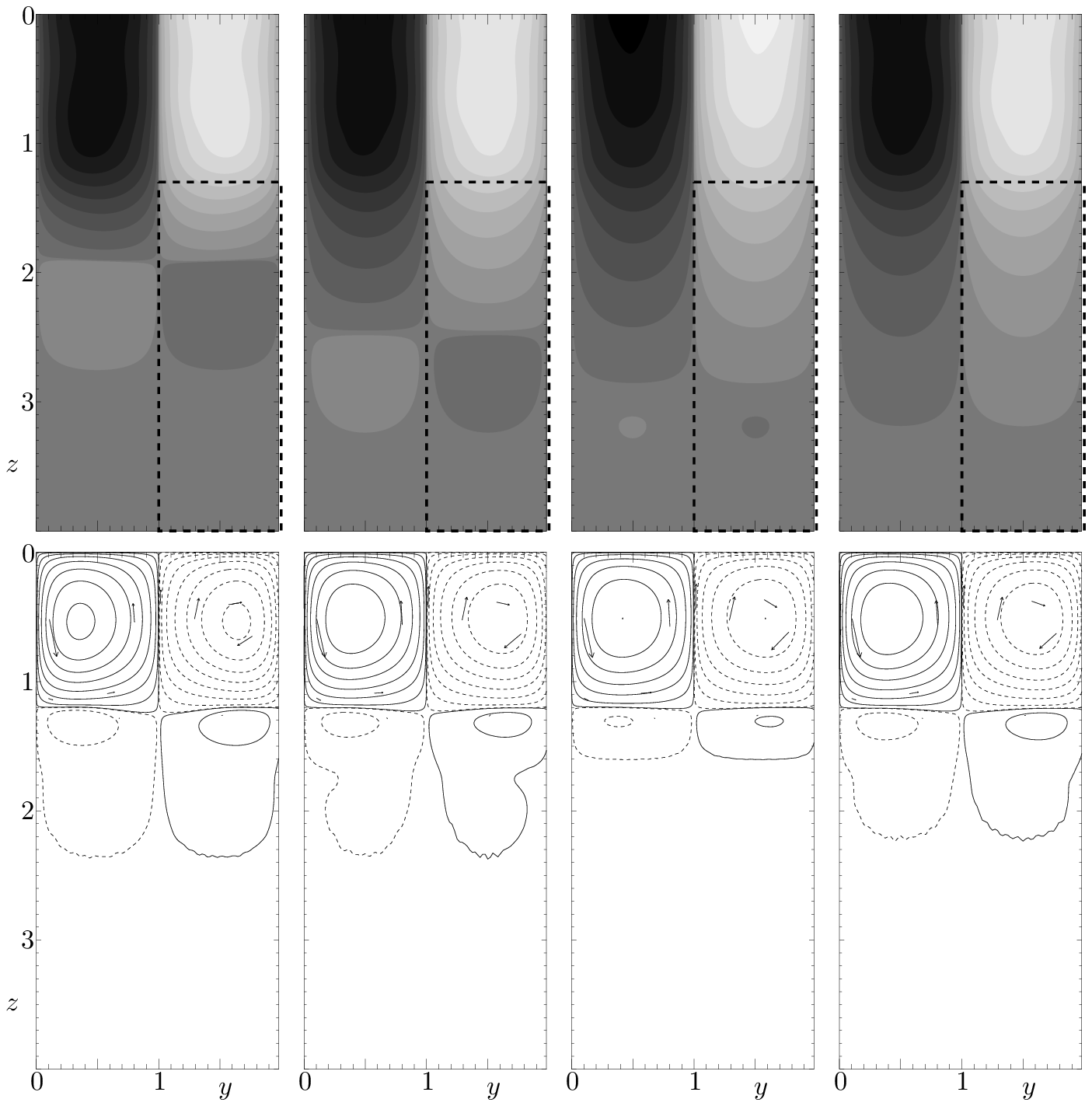}%
  \caption{%
    Shear and meridional flow from Case 2,
    averaged over one rotation period for times $t>t_0$.
    Time averages were taken over the intervals indicated in \fig\ref{fig:high-nu-mom}.
    Contour levels are the same as in \figs\ref{fig:SZ-lam-ave} and \ref{fig:SZ-ave}.}
  \label{fig:high-nu-ave}
\end{figure}

Because of the increased viscosity,
in this case the upper layer is only weakly convective,
and consequently the mean flows are more clearly defined.
Most of the kinetic energy in the upper, convective layer
is associated with the mean azimuthal shear flow, rather
than with convective motions.
The mean azimuthal shear and meridional flow at successive times
$t>t_0$ are plotted in \fig\ref{fig:high-nu-ave}, using the same contour
levels as in \figs\ref{fig:SZ-lam-ave} and \ref{fig:SZ-ave}.
We find that the convection zone's azimuthal shear propagates progressively
deeper into the radiation zone,
as in Cases~0 and 1,
but that meridional flows
within the radiation zone
remain confined
to a thin layer below the bottom of the convection zone.
The convection zone's meridional circulation does not burrow significantly,
and instead a weak counter-circulating meridional cell
is established at the top of the radiation zone
(\cf\ \figs\ref{fig:SZ-lam-ave} and \ref{fig:SZ-ave}).

\begin{figure}[h!]
  \centering%
  \includegraphics[width=13cm]{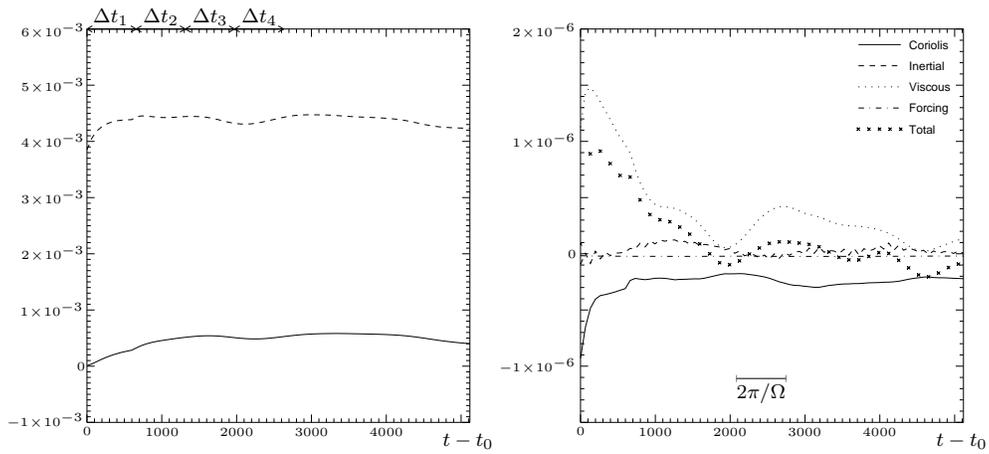}%
  \caption{%
    Same plots as \figs\ref{fig:SZ-mom-lam} and \ref{fig:SZ-mom}, but for Case 2.
    The plots have been averaged over one rotation period.
    In this case the viscous term is dominant, and the Coriolis term takes the
    opposite sign.
    The total duration plotted corresponds to about one viscous diffusion time.}
  \label{fig:high-nu-mom}
\end{figure}

\begin{figure}[h!]
  \centering%
  \includegraphics[height=6cm]{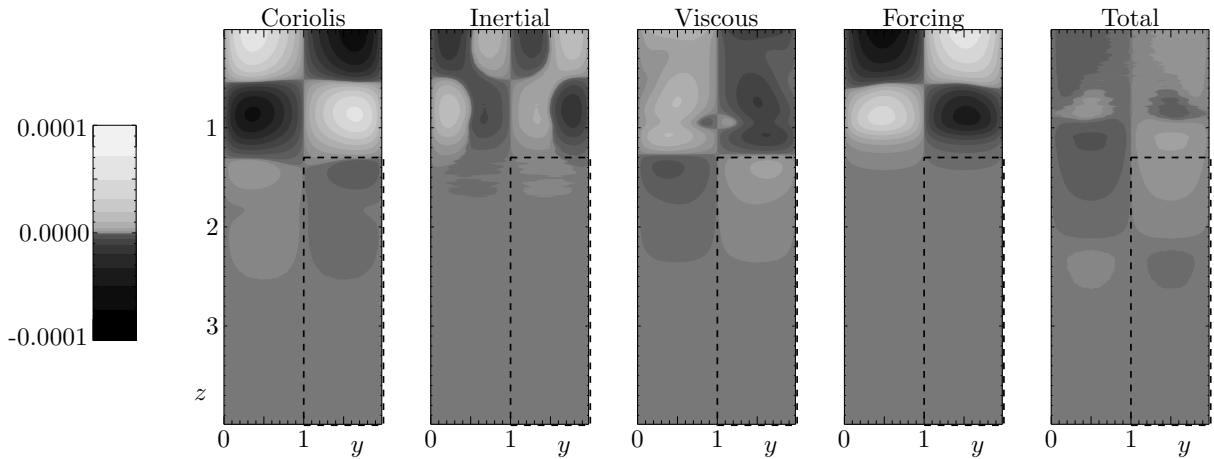}%
  \caption{%
    Same plots as \figs\ref{fig:sasha-SZ-lam} and \ref{fig:sasha-SZ}, but for Case 2,
    and averaged over the time interval $\Delta t_1$ indicated in \fig\ref{fig:high-nu-mom}.
    In this case the viscous term causes the convection zone's azimuthal shear to propagate
    into the radiation zone.  The Coriolis term has the opposite sign to the viscous term,
    and a smaller amplitude overall.}
  \label{fig:sasha-high-nu}
\end{figure}

\Fig\ref{fig:high-nu-mom} shows the equivalent of \figs\ref{fig:SZ-mom-lam}
and \ref{fig:SZ-mom} for Case 2.
In this case,
we find that the propagation of
shear into the radiation zone is caused primarily by viscous diffusion.
After about one viscous diffusion time, a roughly steady state is achieved in which
the viscous force is balanced by the Coriolis force.
We note that the Coriolis force in this case has the opposite sign than in Cases~0
and 1, so
transport of azimuthal momentum
by the mean meridional flow acts to \emph{oppose}
(but not prevent)
the propagation
of the convection zone's shear into the radiation zone.
This is a consequence of the counter-circulating
meridional cell visible in \fig\ref{fig:high-nu-ave},
as demonstrated by
\fig\ref{fig:sasha-high-nu}, which shows vertical cross-sections equivalent to those
of \figs\ref{fig:sasha-SZ-lam} and \ref{fig:sasha-SZ} for Case 2.
The lack of burrowing of the meridional circulation in this case is consistent with the predictions
of laminar models with $\sigma > 1$, as discussed in Section~\ref{sec:parameters}.

\subsection{Case 3: The High-Sigma Regime; Low Thermal Conductivity}

Case 3 has the same values of $\Omega$, $N$, and $\sigma$ as Case 2,
but the viscosity and thermal conductivity are both smaller than in Case 2.
As a result the upper layer is more strongly convective,
with an rms Reynolds number $\simeq60$.
Although this Reynolds number is significantly lower than that of Case 1,
we find that internal waves are generated with a larger amplitude in this case
than in Case 1.
We attribute this
to the lower value of the thermal conductivity in this case,
which reduces the damping of internal waves.
In fact, we find that a standing gravity mode is excited in the lower layer,
which slowly grows in amplitude during the simulation.
The existence of this mode is dependent
on the artificial lower boundary of the computational domain,
and so we end the simulation at the point
where the amplitude of this mode becomes large enough to significantly affect the dynamics
(at around $t = t_0 + 4000$).

\begin{figure}[h!]
  \centering%
  \includegraphics[width=14cm]{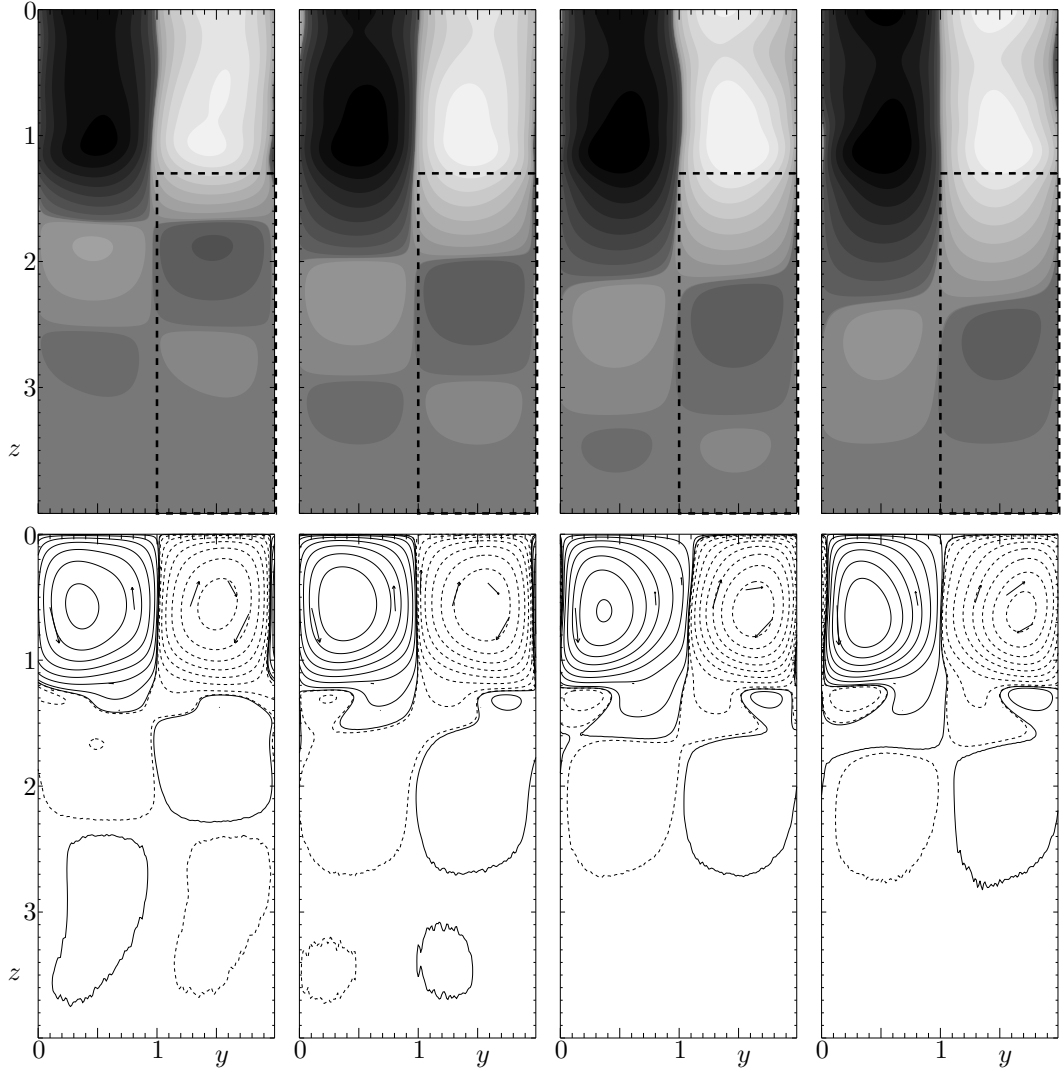}%
  \caption{%
    Shear and meridional flow from Case 3,
    averaged over one rotation period for times $t>t_0$.
    Time averages were taken over the intervals indicated in \fig\ref{fig:low-kappa-mom}.
    Contour levels are the same as in \figs\ref{fig:SZ-lam-ave}, \ref{fig:SZ-ave},
    and \ref{fig:high-nu-ave}.}
  \label{fig:low-k-ave}
\end{figure}

\Fig\ref{fig:low-k-ave} shows the mean azimuthal shear and meridional flow
at successive times $t > t_0$, using the same contour values as
\figs\ref{fig:SZ-lam-ave}, \ref{fig:SZ-ave}, and \ref{fig:high-nu-ave}.
Despite the presence of internal waves, the mean flow exhibits similar behavior to
Case 2.
The meridional circulation driven in the convection zone
turns over at only a short depth within the radiation zone, and a counter-rotating
meridional cell forms beneath, though with a more complicated structure than
in Case 2.  Meanwhile,
the convection zone's shear propagates monotonically into the interior.

\begin{figure}[h!]
  \centering%
  \includegraphics[width=13cm]{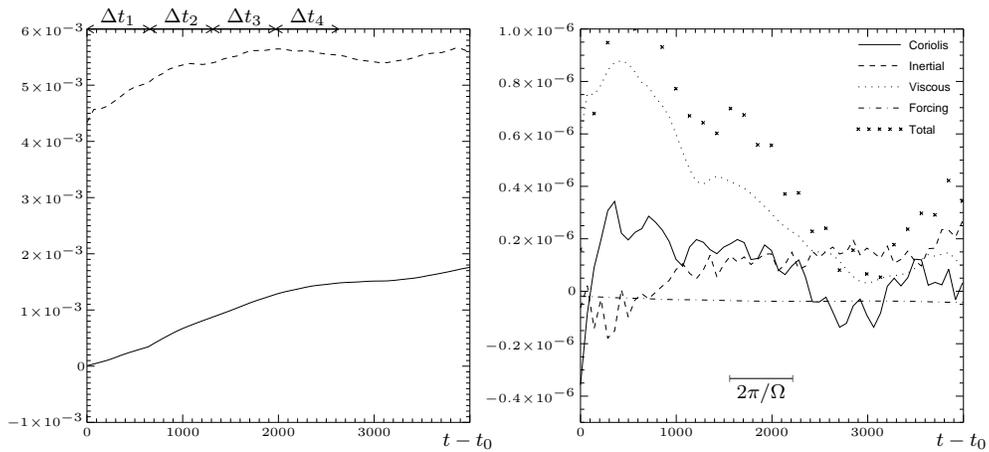}%
  \caption{%
    Same plots as \figs\ref{fig:SZ-mom-lam}, \ref{fig:SZ-mom},
    and \ref{fig:high-nu-mom}, but for Case 3.
    The plots have been averaged over one rotation period.
    The viscous term is dominant, as in Case 2, until $t\simeq t_0+3000$.
    After this time the inertial contribution from a standing gravity mode
    becomes comparable to the viscous term.}
  \label{fig:low-kappa-mom}
\end{figure}

\begin{figure}[h!]
  \centering%
  \includegraphics[height=6cm]{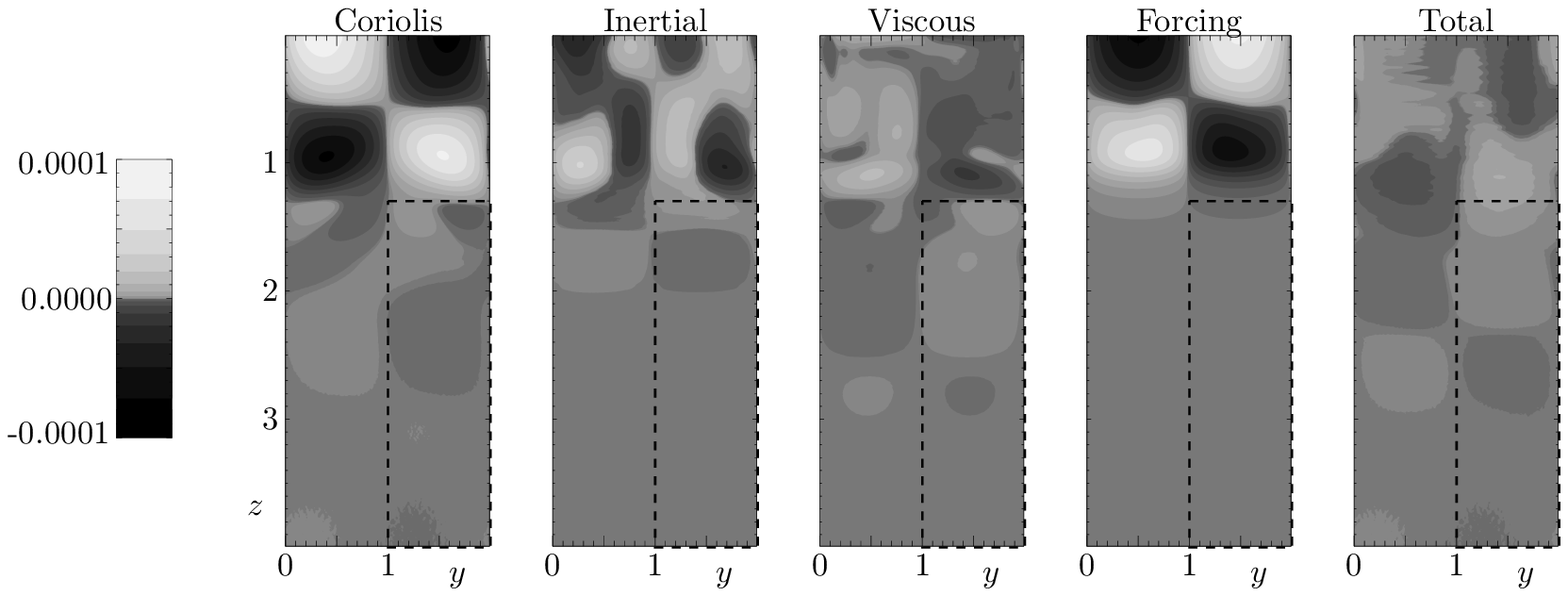}%
  \caption{%
    Same plots as \figs\ref{fig:sasha-SZ-lam}, \ref{fig:sasha-SZ}, and \ref{fig:sasha-high-nu},
    but for Case 3,
    and averaged over the time interval $\Delta t_2$ indicated in \fig\ref{fig:low-kappa-mom}.
    As in Case 2, the viscous term is dominant throughout most of the radiation zone, but
    in this case the Coriolis and inertial terms also contribute to the propagation of the
    convection zone's shear within a thin layer at the top of the radiation zone.}
  \label{fig:sasha-low-kappa}
\end{figure}

\Fig\ref{fig:low-kappa-mom} shows the equivalent of \figs\ref{fig:SZ-mom-lam},
\ref{fig:SZ-mom}, and \ref{fig:high-nu-mom} for Case 3.
As in Case 2,
we find that the propagation of
shear into the radiation zone is caused primarily by viscous diffusion,
although the relative contributions from inertial and Coriolis forces are larger
than in Case 2.
In contrast to Case 2, however, the Coriolis term has the same sign as the viscous term,
at least for times $t \lesssim 2500$.
This is because the convection zone's meridional circulation, shown in \fig\ref{fig:low-k-ave},
manages to burrow a short distance into the radiation zone, as illustrated by
\fig\ref{fig:sasha-low-kappa}, which shows the equivalent of \figs\ref{fig:sasha-SZ-lam},
\ref{fig:sasha-SZ}, and \ref{fig:sasha-high-nu} for Case 3, averaged over the time interval
$\Delta t_2$ indicated in \fig\ref{fig:low-kappa-mom}.
A similar result was obtained by \citet{GaraudAA09} in a steady-state model of the solar
interior; they found that, when $\sigma>1$, meridional circulations that are gyroscopically pumped
within the convection zone extend a distance of order $W/\sigma$ into the radiation zone,
where $W$ is the horizontal lengthscale of the circulation, which in our model is of order unity.

\section{\uppercase{Summary and conclusions}}
\label{sec:summary}

This paper presents the first 3D, self-consistent, and nonlinear study of meridional
flows in the parameter regime described by \citet{Clark73}, which is the relevant
parameter regime for the radiative zones of many solar-type stars.
In this regime the Eddington--Sweet time
is shorter than the viscous time;
the ratio of these timescales is determined by the dimensionless parameter
$\sigma$ defined in \eq(\ref{eq:sigma}).
We have considered four separate cases: two in the ``low-sigma'' regime
and two in the ``high-sigma'' regime.
Our results indicate that the underlying long-time dynamical picture
of angular momentum transport
predicted by laminar models
applies even under
more realistic conditions,
including the presence of overshooting from
the neighboring turbulent convection zone, and the internal waves that this generates.

In Cases~0 and 1, which have $\sigma < 1$,
angular momentum transport is dominated
in the long term by advection
by meridional flows,
and the meridional circulation driven in the convection
zone burrows
(\ie\ extends progressively downward)
into the radiation zone on the Eddington--Sweet timescale,
carrying the differential rotation of the convection zone with it.
The burrowing is more irregular in Case 1 than in Case 0,
as a result of angular momentum transport by turbulence and internal waves.
In Cases~2 and 3, which have $\sigma > 1$,
viscous stresses dominate the transport of angular momentum in the
radiation zone, although there is also a significant contribution from internal
waves in Case 3.  In these two cases the meridional circulation driven
in the convection zone extends only a short distance into the radiation
zone.  The differential rotation of the convection zone still propagates
into the radiation zone, but by viscous diffusion rather than by
meridional advection.

It should be borne in mind that
the simulations presented here are at most weakly turbulent,
in comparison with real stellar convection.
It may be that under the more turbulent conditions characteristic of
real stellar interiors angular momentum transport is dominated by
shear-driven turbulence or internal wave breaking, as argued for instance
by \citet{Zahn92}.
If that is the case then our results may not be applicable to real stars.
However,
our results should certainly be applicable to
previous global-scale
simulations of the solar interior, including those of
\citet{Rogers11}, \citet{Brun-etal11}, and \citet{Strugarek-etal11}.
We note that these simulations were all performed in the ``high-sigma'' regime
in which we found that angular momentum transport is dominated by viscous
stresses.
All of these global models have $\sigma\simeq200$ close to the
top of the radiation zone (although in the model of \citet{Strugarek-etal11}
$\sigma$ drops to around to 20 deeper within the radiation zone).
In each of these
simulations it was indeed found that
viscous stresses
contribute at leading order to the transport of angular momentum within
the radiation zone.  The pattern of meridional flows found in these simulations
is also rather similar to that which we observe in our simulations with $\sigma>1$
(\figs\ref{fig:high-nu-ave} and \ref{fig:low-k-ave}), and no burrowing of the
circulation was observed.  In this situation we expect the transport of angular
momentum between the convection and radiation zones to occur on a viscous
timescale.  This is indeed what \citet{Brun-etal11} and \citet{Strugarek-etal11}
observe.  \citet{Rogers11} reports that an apparently steady state is
achieved in which viscous and Coriolis forces balance, and uniform rotation is
preserved within the radiation zone.  Based on our results,
we suggest that this ``steady state'' is actually evolving slowly on a viscous
timescale.  We are now conducting global simulations in the low-sigma regime
for comparison, to be presented in a forthcoming paper.
Using a global model will also
allow a more realistic study of the effects of internal waves,
avoiding the difficulties encountered in Case 3.

All of the simulations presented here have the same rotation rate and stratification
profile, and in each case the Prandtl number in the radiation zone is smaller than unity.
Yet the strength and depth of the mean meridional circulations vary drastically
between the four cases.  This highlights the danger in
modeling stellar interiors with numerical simulations that have, for example,
realistic rotation and stratification but unrealistic diffusivities.
Our results suggest that the \emph{ordering} of dynamical timescales is of greater
importance than the \emph{exact values} of those timescales when considering
angular momentum transport.
Real stellar interiors are characterized by the same ordering
as the simulations presented here,
but with substantially greater separation.

If our results carry over to realistic stellar parameters then
they have significant implications not only for angular momentum
transport, but also for the transport of chemical elements and magnetic flux by
meridional flows.  In particular, the depth to which chemical elements are carried
from the convection zone into the radiation zone will depend on the depth to
which meridional circulations are able to burrow, and hence on the value of $\sigma$.

An important issue not addressed in the present work is the contribution
of magnetic fields to angular momentum transport in stellar interiors.
If magnetic fields act to suppress differential rotation in the radiation zone, then the burrowing
of meridional flows will also be suppressed \citep{GoughMcIntyre98}.
In that case the role of the magnetic field is analogous to that of the forcing
in the radiation zone in our model for $t < t_0$.
Effects of magnetic fields will be addressed in future papers.

We thank
Pascale Garaud,
Gary Glatzmaier,
C\'eline Guervilly,
Michael McIntyre,
and an anonymous referee
for useful comments and suggestions.
T.S.W.~was supported by NSF CAREER grant 0847477.
N.H.B.~was supported by NASA grant NNX07AL74G
and the Center for Momentum Transport and Flow Organization (CMTFO),
a DoE Plasma Science Center.
Numerical simulations were performed on NSF TeraGrid/XSEDE resources Kraken and Ranger,
and the Pleiades supercomputer at University of California Santa Cruz purchased under
NSF MRI grant AST-0521566.

\end{document}